# AN OVERVIEW ON CLUSTERING METHODS

T. Soni Madhulatha

*Associate Professor, Alluri Institute of Management Sciences, Warangal.*

**ABSTRACT**
Clustering is a common technique for statistical data analysis, which is used in many fields, including machine learning, data mining, pattern recognition, image analysis and bioinformatics. Clustering is the process of grouping similar objects into different groups, or more precisely, the partitioning of a data set into subsets, so that the data in each subset according to some defined distance measure. This paper covers about clustering algorithms, benefits and its applications. Paper concludes by discussing some limitations.

*Keywords*: Clustering, hierarchical algorithm, partitional algorithm, distance measure,

## I. INTRODUCTION

Clustering can be considered the most important unsupervised learning problem; so, as every other problem of this kind, it deals with finding a structure in a collection of unlabeled data. A cluster is therefore a collection of objects which are "similar" between them and are "dissimilar" to the objects belonging to other clusters. Besides the term data clustering as synonyms like cluster analysis, automatic classification, numerical taxonomy, botrology and typological analysis.

## II. TYPES OF CLUSTERING.

Data clustering algorithms can be hierarchical or partitional. Hierarchical algorithms find successive clusters using previously established clusters, whereas partitional algorithms determine all clusters at time. Hierarchical algorithms can be agglomerative (bottom-up) or divisive (top-down). Agglomerative algorithms begin with each element as a separate cluster and merge them in successively larger clusters. Divisive algorithms begin with the whole set and proceed to divide it into successively smaller clusters.

## HIERARCHICAL CLUSTERING

A key step in a hierarchical clustering is to select a distance measure. A simple measure is manhattan distance, equal to the sum of absolute distances for each variable. The name comes from the fact that in a two-variable case, the variables can be plotted on a grid that can be compared to city streets, and the distance between two points is the number of blocks a person would walk.

A more common measure is Euclidean distance, computed by finding the square of the distance between each variable, summing the squares, and finding the square root of that sum. In the two-variable case, the distance is analogous to finding the length of the hypotenuse in a triangle; that is, it is the distance "as the crow flies." A review of cluster analysis in health psychology research found that the most common distance measure in published studies in that research area is the Euclidean distance or the squared Euclidean distance.

The **Manhattan distance function** computes the distance that would be traveled to get from one data point to the other if a grid-like path is followed. The Manhattan distance between two items is the sum of the differences of their corresponding components. The formula for this distance between a point X=(X1, X2, etc.) and a point Y=(Y1, Y2, etc.) is:

$$d = \sum_{i=1}^{n} |X_i - Y_i|$$

Where n is the number of variables, and *Xi* and *Yi* are the values of the *i*th variable, at points *X* and *Y* respectively.

The **Euclidean distance function** measures the 'as-the-crow-flies distance. The formula for this distance between a point *X* (*X1, X2,* etc.) and a point *Y* (*Y1, Y2,* etc.) is:

$$d = \sqrt{\sum_{j=1}^{n}(x_j - y_j)^2}$$

Deriving the Euclidean distance between two data points involves computing the square root of the sum of the squares of the differences between corresponding values.

The following figure illustrates the difference between Manhattan distance and Euclidean distance:

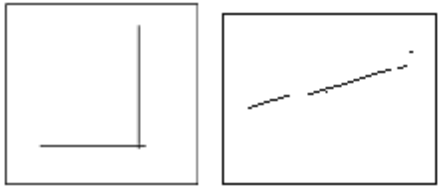

Manhattan distance        Euclidean distance

This method builds the hierarchy from the individual elements by progressively merging clusters. Again, we have six elements {a} {b} {c} {d} {e} and {f}. The first step is to determine which elements to merge in a cluster. Usually, we want to take the two closest elements, therefore we must define a distance between elements. One can also construct a distance matrix at this stage.

For example, suppose these data are to be analyzed, where pixel euclidean distance is the distance metric.

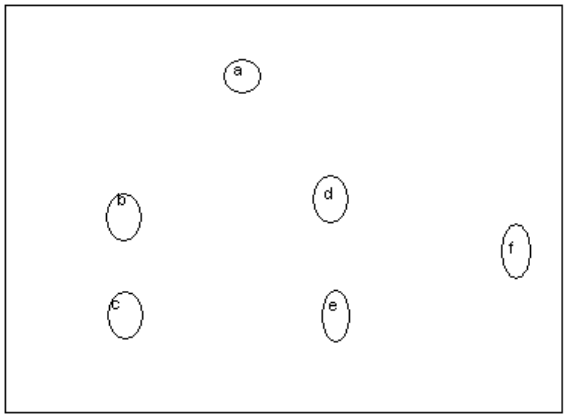

Usually the distance between two clusters and is one of the following:

- The maximum distance between elements of each cluster is also called complete linkage clustering.

$$\max \{d(x, y) : x \in A, y \in B\}$$

- The minimum distance between elements of each cluster is also called single linkage clustering.

$$\min \{d(x, y) : x \in A, y \in B\}$$

- The mean distance between elements of each cluster is also called average linkage clustering.

$$\frac{1}{card(A) card(B)} \sum_{x \in A} \sum_{y \in B} d(x, y)$$

- The sum of all intra-cluster variance
- The increase in variance for the cluster being merged
- The probability that candidate clusters spawn from the same distribution function.

Each agglomeration occurs at a greater distance between clusters than the previous agglomeration, and one can decide to stop clustering either when the clusters are too far apart to be merged or when there is a sufficiently small number of clusters.

Agglomerative hierarchical clustering

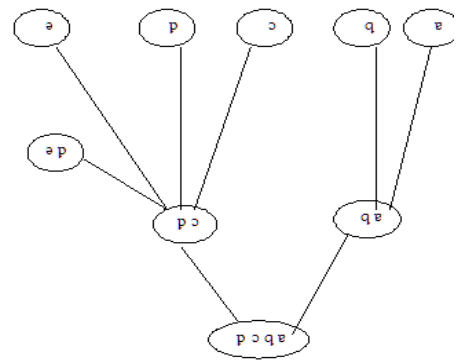

**Divisive clustering**
So far we have only looked at agglomerative clustering, but a cluster hierarchy can also be generated top-down. This variant of hierarchical clustering is called top-down clustering or divisive clustering. We start at the top with all documents in one cluster. The cluster is split using a flat clustering algorithm. This procedure is applied recursively until each document is in its own singleton cluster.

Top-down clustering is conceptually more complex than bottom-up clustering since we need a second, flat clustering algorithm as a ``subroutine''. It has the advantage of being more efficient if we do not generate a complete hierarchy all the way down to individual document leaves. For a fixed number of top levels, using an efficient flat algorithm like K-means, top-down algorithms are linear in the number of documents and clusters

Hierarchal method suffers from the fact that once the merge/split is done, it can never be undone. This rigidity is useful in that is useful in that it leads to smaller computation costs by not worrying about a combinatorial number of different choices.

However there are two approaches to improve the quality of hierarchical clustering

Perform careful analysis of object linkages at each hierarchical partitioning such as CURE and Chameleon.

Integrate hierarchical agglomeration and then redefine the result using iterative relocation as in BRICH

**PARTITIONAL CLUSTERING:**
Partitioning algorithms are based on specifying an initial number of groups, and iteratively reallocating objects among groups to convergence. This algorithm typically determines all clusters at once. Most applications adopt one of two popular heuristic methods like

   k-mean algorithm

   k-medoids algorithm

**K-means algorithm**
The K-means algorithm assigns each point to the cluster whose center also called centroid is nearest. The center is the average of all the points in the cluster that is, its coordinates are the arithmetic mean for each dimension separately over all the points in the cluster.

The pseudo code of the k-means algorithm is to explain how it works:
A. Choose K as the number of clusters.
B. Initialize the codebook vectors of the K clusters (randomly, for instance)
C. For every new sample vector:
 a. Compute the distance between the new vector and every cluster's codebook vector.
 b. Re-compute the closest codebook vector with the new vector, using a learning rate that decreases in time.

The reason behind choosing the k-means algorithm to study is its popularity for the following reasons:
- Its time complexity is O (nkl), where n is the number of patterns, k is the number of clusters, and l is the number of iterations taken by the algorithm to converge.
- Its space complexity is O (k+n). It requires additional space to store the data matrix.
- It is order-independent; for a given initial seed set of cluster centers, it generates the same partition of the data irrespective of the order in which the patterns are presented to the algorithm.

**K-medoids algorithm:**
The basic strategy of k-medoids algorithm is each cluster is represented by one of the objects located near the center of the cluster. PAM (Partitioning around Medoids) was one of the first k-medoids algorithm is introduced.

 The pseudo code of the k-medoids algorithm is to explain how it works:

Arbitrarily choose k objects as the initial medoids Repeat Assign each remaining object to the cluster with the nearest medoids Randomly select a non-medoid object $O_{random}$ Compute the total cost, S, of swapping $O_j$ with $O_{random}$

 If S<0 the swap $O_j$ with $O_{random}$ to form the new set of k-medoids
 Until no changes

K-medoids method is more robust than k-mean in presence of noise and outliers because a medoids is less influenced by outliers or other extreme values than a mean.

**DENSITY-BASED CLUSTERING**
Density-based clustering algorithms are devised to discover arbitrary-shaped clusters. In this approach, a cluster is regarded as a region in which the density of data objects exceeds a threshold. DBSCAN and SSN are two typical algorithms of this kind.

**DBSCAN algorithm**
The DBSCAN algorithm was first introduced by Ester, and relies on a density-based notion of clusters. Clusters are identified by looking at the density of points. Regions with a high density of points depict the existence of clusters whereas regions with a low density of points indicate clusters of noise or clusters of outliers. This algorithm is particularly suited to deal with large datasets, with noise, and is able to identify clusters with different sizes and shapes.

The key idea of the DBSCAN algorithm is that, for each point of a cluster, the neighbourhood of a given radius has to contain at least a minimum number of points, that is, the density in the neighbourhood has to exceed some predefined threshold.

This algorithm needs three input parameters:

- k, the neighbour list size;

- Eps, the radius that delimitate the neighbourhood area of a point (Eps neighbourhood);

- MinPts, the minimum number of points that must exist in the Eps-neighbourhood.

The clustering process is based on the classification of the points in the dataset as core points, border points and noise points, and on the use of density relations between points to form the clusters.

The pseudo code of the DBSCAN algorithm is to explain how it works:

To clusters a dataset, our DBSCAN implementation starts by identifying the k nearest neighbours of each point and identify the farthest k nearest neighbour. The average of all this distance is then calculated.

For each point of the dataset the algorithm identifies the directly density-reachable points using the Eps threshold provided by the user and classifies the points into core or border points.

It then loop trough all points of the dataset and for the core points it starts to construct a new cluster with the support of the GetDRPoints() procedure that follows the definition of density reachable points.

In this phase the value used as Eps threshold is the average distance calculated previously. At the end, the composition of the clusters is verified in order to check if there exist clusters that can be merged together. This can append if two points of different clusters are at a distance less than Eps.

Note: DBSCAN does not deal very well with clusters of different densities.

## SNN ALGORITHM

The SNN algorithm, as DBSCAN, is a density-based clustering algorithm. The main difference between this algorithm and DBSCAN is that it defines the similarity between points by looking at the number of nearest neighbours that two points share. Using this similarity measure in the SNN algorithm, the density is defined as the sum of the similarities of the nearest neighbours of a point. Points with high density become core points, while points with low density represent noise points. All remainder points that are strongly similar to a specific core points will represent a new clusters.

The SNN algorithm needs three inputs parameters:

- K, the neighbours' list size;

- Eps, the threshold density;

- MinPts, the threshold that define the core points.

The pseudo code of the SSN algorithm is to explain how it works:

Define the input parameters.

Find the K nearest neighbours of each point of the dataset.

Then the similarity between pairs of points is calculated in terms of how many nearest neighbours the two points share. Using this similarity measure, the density of each point can be calculated as being the numbers of neighbours with which the number of shared neighbours is equal or greater than Eps.

The points are classified as being core points, if the density of the point is equal or greater than MinPts. At this point, the algorithm has all the information needed to start to build the clusters. Those start to be constructed around the core points.

However, these clusters do not contain all points. They contain only points that come from regions of relatively uniform density. The points that are not classified into any cluster are classified as noise points.

## GRID-BASED CLUSTERING

The grid based clustering approach uses a multiresolution grid data structure. It quantizes the space into a finite number of cells that form a grid structure on which all the operations for clustering are performed. Grid approach includes STING (STatistical INformation Grid) approach and CLIQUE

**Basic Grid-based Algorithm**
1. Define a set of grid-cells
2. Assign objects to the appropriate grid cell and compute the density of each cell.
3. Eliminate cells, whose density is below a certain threshold t.
4. Form clusters from contiguous (adjacent) groups of dense cells.

The pseudo code of the STING algorithm is to explain how it works:

The spatial area is divided into rectangular cells
There are several levels of cells corresponding to different levels of resolution

Each cell is partitioned into a number of smaller cells in the next level. Statistical info of each cell is calculated and stored beforehand and is used to answer queries

Parameters of higher level cells can be easily calculated from parameters of lower level cell count, mean, s, min, max type of distribution—normal, uniform, etc.
Use a top-down approach to answer spatial data queries

Start from a pre-selected layer—typically with a small number of cells from the pre-selected layer until you reach the bottom layer do the following:

For each cell in the current level compute the confidence interval indicating a cell's relevance to a given query;
1.  If it is relevant, include the cell in a cluster

2. If it irrelevant, remove cell from further consideration

3. otherwise, look for relevant cells at the next lower layer

1. Combine relevant cells into relevant regions (based on grid-neighborhood) and return the so obtained clusters as your answers.

**Advantages:**
Query-independent, easy to parallelize, incremental update $O(K)$, where $K$ is the number of grid cells at the lowest level

**Disadvantages:**
All the cluster boundaries are either horizontal or vertical, and no diagonal boundary is detected

**MODEL-BASED CLUSTERING**
Model-Based Clustering methods attempt to optimize the fit between the given data and some mathematical model. Such methods often based on the assumption that the data are generated by mixture of underlying probability distributions. Model-Based Clustering methods follow two major approaches: Statistical Approach or Neural network approach

1. Clustering is also performed by having several units competing for the current object
2. The unit whose weight vector is closest to the current object wins
3. The winner and its neighbors learn by having their weights adjusted
4. SOMs are believed to resemble processing that can occur in the brain
5. Useful for visualizing high-dimensional data in 2- or 3-D space

In model-based clustering, the data x are viewed as coming P from a mixture density

$$f(x) = \sum_{k=1}^{G} T_k f_k(x)$$

where $f_k$ is the probability density function of the observations in group k, and $T_k$ is the probability that an observation comes from the kth mixture component

Each component is usually modeled by the normal or Gaussian distribution. Component distributions are characterized by the mean $\mu_k$ and the covariance matrix $\sum_k$, and have the probability density function

$$\phi(x_i; \mu_k, \sum{}_k) = \frac{\exp\left\{-\frac{1}{2}(x_i - \mu_k)^T \sum{}_k^{-1} (x_i - \mu_k)\right\}}{\sqrt{\det(2\prod \sum{}_k)}}$$

For univariate data, the covariance matrix reduces to a scalar variance. The likelihood for data consisting of n observations assuming a Gaussian mixture model with G multivariate mixture components is

$$\prod_{i=1}^{n} \sum_{k=1}^{G} T_k \phi(x_i; \mu_k, \sum{}_k).$$

MCLUST is probably the most well known model-based

This is all about various clustering algorithms.

**III. HOW TO DETERMINE THE NUMBER OF CLUSTERS**
Many clustering algorithms require the specification of the number of clusters to produce in the input data set, prior to execution of the algorithm. Barring knowledge of the proper value beforehand, the appropriate value must be determined, a problem on its own for which a number of techniques have been developed.

− If the number of clusters known, termination condition is given!
− In general, set a distance threshold value (termination condition)
− The *K*-cluster lifetime as the range of threshold values on the dendrogram tree that leads to the identification of *K* clusters
− Heuristic rule: cut a dendrogram tree with maximum life time

One simple rule of thumb sets the number to

$k \approx \sqrt{n/2}$  with *n* as the number of objects .

**Elbow criterion**
The elbow criterion is a common rule of thumb to determine what number of clusters should be chosen, for example for *k*-means and agglomerative hierarchical clustering. The elbow criterion says that you should choose a number of clusters so that adding another cluster doesn't add sufficient information. More precisely, if you graph the percentage of variance explained by the clusters against the number of clusters, the first clusters will add much information, but at some point the marginal gain will drop, giving an angle in the graph.

Another set of methods for determining the number of clusters are information criteria, such as :

The Akaike information criterion (AIC), The Bayesian information criterion (BIC), The Deviance information criterion (DIC).

## IV. HOW ALGORITHMS ARE COMPARED

The above clustering algorithms are compared according to the following factors:
The size of the dataset,
Number of the clusters,
Type of dataset,
Type of software

Table 1 explains how the four algorithms are compared and the conclusions are written down.

| | Size of Dataset | Number of Clusters | Type of Dataset | Type of Software |
|---|---|---|---|---|
| Partitional Algorithm | Huge Dataset & Small Dataset | Large number of clusters & Small number of clusters | Ideal Dataset & Random Dataset | LNKnet Package & Cluster and TreeView Package |
| Hierarchical Algorithm | Huge Dataset & Small Dataset | Large number of clusters & Small number of Clusters | Ideal Dataset & Random Dataset | LNKnet Package & Cluster and TreeView Package |
| Grid based Algorithm | Huge Dataset & Small Dataset | Large number of clusters & Small number of clusters | Ideal Dataset & Random Dataset | LNKnet Package & Cluster and TreeView Package |
| Model-based Algorithm | Huge Dataset & Small Dataset | Large number of clusters & Small number of clusters | Ideal Dataset & Random Dataset | LNKnet Package & Cluster and TreeView Package |

## V. POSSIBLE APPLICATIONS

Clustering algorithms can be applied in many fields, for instance:

- Marketing: finding groups of customers with similar behavior given a large database of customer data containing their properties and past buying records;
- Financial task: Forecasting stock market, currency exchange rate, bank bankruptcies, un-derstanding and managing financial risk, trading futures, credit rating,
- Biology: classification of plants and animals given their features;
- Libraries: book ordering;
- Insurance: identifying groups of motor insurance policy holders with a high average claim cost; identifying frauds;
- City-planning: identifying groups of houses according to their house type, value and geographical location;
- Earthquake studies: clustering observed earthquake epicenters to identify dangerous zones;
- WWW: document classification; clustering web log data to discover groups of similar access patterns

## VI. CONCLUSION

Clustering is a descriptive technique. The solution is not unique and it strongly depends upon the analyst's choices. We described how it is possible to combine different results in order to obtain stable clusters, not depending too much on the criteria selected to analyze data. Clustering always provides groups, even if there is no group structure.
When applying a cluster analysis we are hypothesizing that the groups exist. But this assumption may be false or weak. Clustering results' should not be generalized**.** Cases in the same cluster are similar only with respect to the information cluster analysis was based on i.e., dimensions/variables inducing the dissimilarities.

## REFERENCES
1. Han, J. and Kamber, M. Data Mining: Concepts and Techniques, 2001 (Academic Press, San Diego, California, USA).
2. Comparision between clustering algorithms- Osama Abu Abbas.
3. Pham, D.T. and Afify, A.A. Clustering techniques and their applications in engineering. Submitted to Proceedings of the Institution of Mechanical Engineers,